\documentclass[11pt]{revtex4}

\usepackage{graphicx}

\begin{document}

\title{A Macroscopic Classical System with Entanglement}

\author{D.W. Snoke}

\affiliation{Department of Physics and Astronomy, University of Pittsburgh\\3941 O'Hara St., Pittsburgh, PA 15260}

\begin{abstract}
It is possible to construct a classical, macroscopic system which has a mathematical structure that is exactly the same as that of a quantum mechanical system and which can be put into a state which is identical to quantum mechanical entanglement. This paper presents a simple example, including a way in which the system can be measured to violate Bell's inequalities. This classical simulation of a quantum system allows us to visualize entanglement and also helps us to see what aspects of quantum mechanical systems are truly nonclassical. 
\end{abstract}

\maketitle

It is sometimes argued that entanglement is a uniquely quantum mechanical property which cannot occur in classical systems. This is incorrect, although the degree of entanglement in quantum mechanical systems has no upper bound, while in classical systems there is an upper bound given by the dimensionality of space. 

The canonical case of entanglement in quantum mechanics is given by a superposition of the form
\begin{equation}
|\Psi\rangle = \frac{1}{\sqrt{2}}(|0\rangle|1\rangle +i|1\rangle|0\rangle ),
\label{entangle}
\end{equation}
where $|0\rangle$ and $|1\rangle$ are two states available to two different quantum mechanical subsystems. Such a state is not factorizable into a product of states in each subsystem. 

This state is physically realized, for example, in the case of a beamsplitter which has one photon impinging on it. In this case $|0\rangle$ corresponds to one output of the beamsplitter having no photon, and $|1\rangle$ corresponds to the output having one photon. The product state gives the total state of both outputs of the beamsplitter. This state is the result of the standard 50-50 beamsplitter matrix operator \cite{mandel}
\begin{equation}
M = 
\frac{1}{\sqrt{2}}
\left(
\begin{array}{cc}1 & i\\
i & 1 
\end{array}
\right)
\end{equation}
acting on the input state $|1\rangle|0\rangle$, which is written in vector form as $(1,0)$, and corresponds to one photon entering the beamsplitter from one direction. 

To see how to simulate the state (\ref{entangle}) classically, we must begin by recalling how photon operators and states are defined. Photons are defined as the eigenstates of the Hamiltonian 
\begin{equation}
H = \sum_k \hbar\omega_k \left(\hat{N}_k+\frac{1}{2}\right) =   \sum_k \hbar\omega_k \left(a^{\dagger}_ka^{ }_k +\frac{1}{2}\right),
\end{equation}
where $a^{\dagger}_k$ and $a^{ }_k$ are the creation and destruction operators for the wave mode $k$, and $\omega_k$ is the frequency of the mode $k$. As shown in many textbooks \cite{snokebook,louisell}, each wave mode $k$ is an independent harmonic oscillator, such that the creation and destruction operators obey the commutation relation
\begin{equation}
[a^{ }_k,a^{\dagger}_k] = 1. 
\end{equation}
This relation follows from the underlying wave equation for the harmonic oscillator,
\begin{equation}
H\psi = i\hbar\frac{\partial \psi}{\partial t} = \left[\frac{p_k^2}{2m} + \frac{1}{2}\gamma x_k^2\right]\psi = \left[-\frac{\hbar^2}{2m}\frac{\partial^2}{\partial x_k^2} + \frac{1}{2}\gamma x_k^2\right]\psi,
\label{oscwave}
\end{equation}
where $\psi$ is a wave function. Here we have used an effective mass $m$ and spring constant $\gamma$, which are appropriate for phonons in a system of coupled atoms, but photons in a vacuum have exactly the same mathematical structure \cite{snokebook}, if we substitute 
\begin{eqnarray}
m/a^3 \rightarrow \epsilon_0, \hspace{.4cm} a/\gamma\rightarrow \mu_0, \hspace{.4cm}
x_k \rightarrow A_k, 
\label{mapping}
\end{eqnarray}
where $a$ is the size of the local oscillator with mass $m$, $\epsilon_0$ and $\mu_0$ are the permittivity and permeability of free space used in Maxwell's equations, and $A_k$ is the vector potential of electromagnetism. Instead of $x_k$ for the spatial displacement of the oscillator $k$, we have the strength of the electromagnetic field $A_k$.  The wave function $\psi$ gives the probability of a given value of $x_k$ or $A_k$. 

In this algebra, the destruction operator is defined as
\begin{eqnarray}
a^{ }_k = \frac{1}{\sqrt{2}}\left(\sqrt{\frac{m\omega_k}{\hbar}}x_k +\frac{i}{\sqrt{m\hbar\omega_k}}p_k \right) = 
\frac{1}{\sqrt{2}}\left(\sqrt{\frac{m\omega_k}{\hbar}}x_k +\sqrt{\frac{\hbar}{m\omega_k}}\frac{\partial}{\partial x_k}\right) = \frac{1}{\sqrt{2}}\left(\tilde{x}_k+\frac{\partial}{\partial\tilde{x}_k}\right),
\end{eqnarray}
where $\tilde{x}_k = (\sqrt{m\omega_k/\hbar})x_k$. Similarly, the creation operator is defined as
\begin{equation}
a^{\dagger}_k = \frac{1}{\sqrt{2}}\left(\tilde{x}_k-\frac{\partial}{\partial\tilde{x}_k}\right).
\end{equation}
With these definitions, it is easy to show that the eigenstates $|0\rangle$ and $|1\rangle$ correspond to
\begin{eqnarray}
\psi_0(x_k) &=& \langle x_k|0\rangle = \frac{1}{{\pi}^{1/4}}e^{-\tilde{x}_k^2/2} \\
\psi_1(x_k) &=& \langle x_k|1\rangle = \frac{\sqrt{2}}{{\pi}^{1/4}}\tilde{x}_ke^{-\tilde{x}_k^2/2},
\label{states}
\end{eqnarray}
with $\omega_k = \sqrt{\gamma/m}$, and $a^{\dagger}_k$ and $a^{ }_k$ have the standard actions $a^{\dagger}_k|0\rangle = |1\rangle$ and $a^{ }_k|1\rangle = |0\rangle$.
 
 
 
Thus, the ground state of the photon mode $k$, corresponding to no photon, is a wave function $\psi$ which is a Gaussian, and the first excited state, corresponding to one photon, is a wave function $\psi$ which is a Gaussian multiplied by $\sqrt{2}x_k$. This wave function $\psi$ is not the same as the electromagnetic field function of the mode $k$ in real space. The electromagnetic field of mode $k$ is given by $A(z,t) = A_ke^{i(kz-\omega_k t)}$; the wave function $\psi(A_k)$, which is the same as $\psi(x_k)$ here, gives the probability of finding a particular amplitude $A_k$. If no measurement is made of $A_k$, however, then $\psi$ is a continuous function which satisfies the wave equation (\ref{oscwave}).

{\bf Cavity resonators with effective mass and spring constant}.
 The question is then whether there is a classical system that obeys the wave equation (\ref{oscwave}). The answer is yes; we can construct a system with this wave equation using a classical optical resonator. 

We imagine a classical resonator comprised of two parallel mirrors separated by a distance $L$. 
The classical Maxwell wave equation which applies in this system is
\begin{equation}
\nabla^2E = \frac{1}{c^2}\frac{\partial^2 E}{\partial t^2}, 
\label{maxwell}
\end{equation}
where we ignore the polarization of the electric field; in all of the following we assume that the electric field is always polarized in one direction.  We write the solution of this wave equation subject to the cavity boundary conditions as
 \begin{equation}
E =\psi \cos(k_{\perp}z)e^{-i\omega t},  
\end{equation} 
where $k_{\perp} = N\pi/L$; only integer values of $N$ are allowed, because the perpendicular component $k_{\perp}$ is quantized by the boundary condition that the electric field must vanish at the surface of the mirrors. The amplitude $\psi$ may vary in time and in space along the plane of the cavity. We write this envelope amplitude suggestively as $\psi$ because we will see that it plays the same role as the harmonic oscillator wave function $\psi$.

Keeping only leading terms in frequency (known as the slowly varying envelope approximation \cite{snokebook,yariv}), we have for the time derivative of $E$,
\begin{eqnarray}
\frac{\partial^2 E}{\partial t^2} 
&\simeq&  \left(-\omega^2\psi -2i\omega\frac{\partial \psi}{\partial t}\right)\cos(k_{\perp}z)e^{-i\omega t},  
\end{eqnarray}
The Maxwell wave equation (\ref{maxwell}) then becomes
\begin{eqnarray}
&& (-k_{\perp}^2\psi + \nabla_{\|}^2\psi)
=\frac{1}{c^2}\left(-\omega^2\psi -2i\omega\frac{\partial \psi}{\partial t}\right)
.
\label{full}
\end{eqnarray}
We allow that $k_{\perp}$ may vary slowly along the plane of the cavity, due to varying cavity thickness $L$. 
In particular, if we arrange to have a maximum of the thickness $L$ at position $x = 0$, with parabolic variation of the thickness away from $x=0$,  we can write
\begin{equation}
k_{\perp}^2 = \frac{N^2\pi^2}{L^2(x)} =   N^2\pi^2 \frac{1}{(L_{0} - b x^2)^2} \simeq  \frac{N^2\pi^2}{L^2_{0}}(1 +2 b x^2/L_0) \equiv \frac{\omega_0^2}{c^2} (1 +2 b x^2/L_0),
\end{equation}
where $b$ is a constant that gives the variation of $L(x)$ in the plane, and $\omega_0 = N\pi/L_0$.  

Picking $\omega \simeq \omega_0$, the Maxwell wave equation (\ref{full}) becomes
\begin{eqnarray}
&& \nabla_{\|}^2\psi -\frac{2\omega_0^2 b}{c^2L_0}x^2\psi
=\frac{1}{c^2}\left(-2i\omega_0\frac{\partial \psi}{\partial t}\right)
\label{opticalGP}
\end{eqnarray}
Rearranging, we have
\begin{equation}
-\frac{c^2}{2\omega_0}\nabla_{\|}^2\psi  + \frac{b\omega_0 }{L_0}x^2\psi = i\frac{\partial \psi}{\partial t}.
\label{equiv}
\end{equation}
This is equivalent to (\ref{oscwave}) if we assign $m =\hbar \omega_0/c^2$ and $\gamma = 2\hbar \omega_0(b/L_0)$.
The solutions of this equation are already well known, namely the solutions of the quantum harmonic oscillator discussed above, with evenly spaced frequencies. 

We have made two assumptions to arrive at this result, namely that the cavity thickness is thin enough that $\omega_0$ is well above the rate of change of the envelope function $\psi$, and the gradient of the cavity thickness is small enough that the cavity can be treated as locally planar. Both of these limits are easily achieved in experiments, and such experiments have been done in at least two cases. One possibility is to vary the index of refraction in a parabolic fashion, giving the equivalent behavior by changing the effective velocity $c$ instead of $L$ in the above. This was invoked in a proposal \cite{heberle,botao} for modelocking of a very small cavity laser using the evenly spaced frequencies for the lateral modes in the plane of the cavity instead of the standard modelocking method of using the evenly spaced longitudinal modes. The time-varying laser mode in this proposal corresponds to two pulses moving in counter-propagating circles in the plane of the cavity, rather than a pulse bouncing back and forth between the two mirrors. This limit has also been used in the recent ``photon condensate" experiments \cite{photonBEC}; the variation of the cavity thickness gave a harmonic potential in the plane which could be used to trap the photons in the ground state at the center of the cavity, which is a Gaussian mode. 

This type of resonator is therefore standard optics, not exotic, and can easily be fabricated for experimental studies using either varying cavity thickness or index of refraction variation. If the optical modes are coupled to electronic transitions, this leads to a nonlinear term which makes (\ref{equiv}) become a standard Gross-Pitaevskii equation \cite{snoke-chapter}, also known as a nonlinear Schr\"odinger equation. This is the basis of the many experiments done with condensates of polaritons \cite{polref}, which are essentially photons dressed with hard-core repulsion, leading the polariton condensates to obey a Gross-Pitaevslii equation \cite{berloff}.


{\bf Entangled states of the resonator}. The fact that the resonator discussed above has two spatial dimensions in the plane allows us to create entangled states exactly equivalent to (\ref{entangle}). Since linear waves obey the principle of superposition, we can make superpositions of macroscopic electromagnetic waves just as we do with quantum mechanical wave functions. The state equivalent to (\ref{entangle}) is
\begin{equation}
\psi(x,y,t) = \frac{1}{{\pi}^{1/2}}e^{-{x}^2/2} {y}e^{-{y}^2/2}e^{i\tilde\omega t} + \frac{i}{{\pi}^{1/2}}{x}e^{-{x}^2/2} e^{-{y}^2/2} e^{i\tilde\omega t},
\label{wave}
\end{equation}
where $\tilde\omega = \sqrt{\gamma/m} = \sqrt{2(b/L_0)}c$. This frequency can be quite low compared to the frequency $\omega_0$ at which the electromagnetic field oscillates, if the curvature of the mirrors is low.

\begin{figure}[htbp]
\begin{center}
\includegraphics[width=0.46\textwidth]{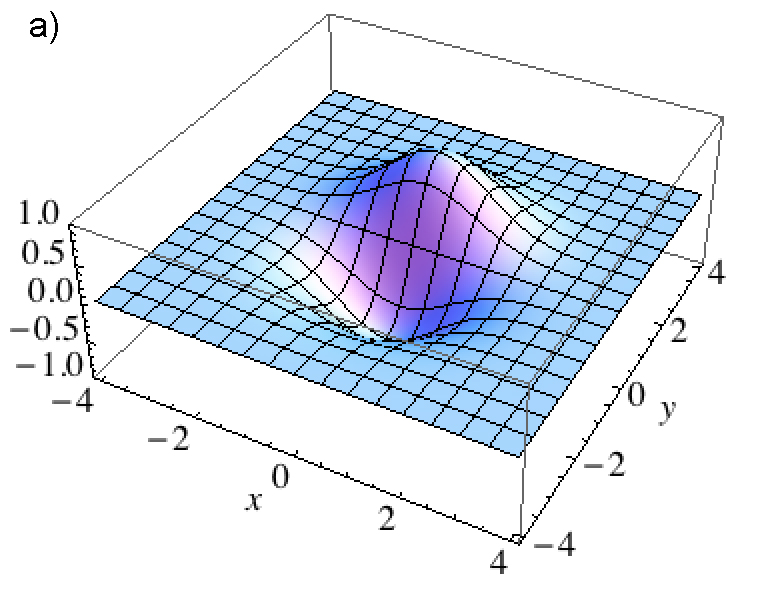}
\includegraphics[width=0.48\textwidth]{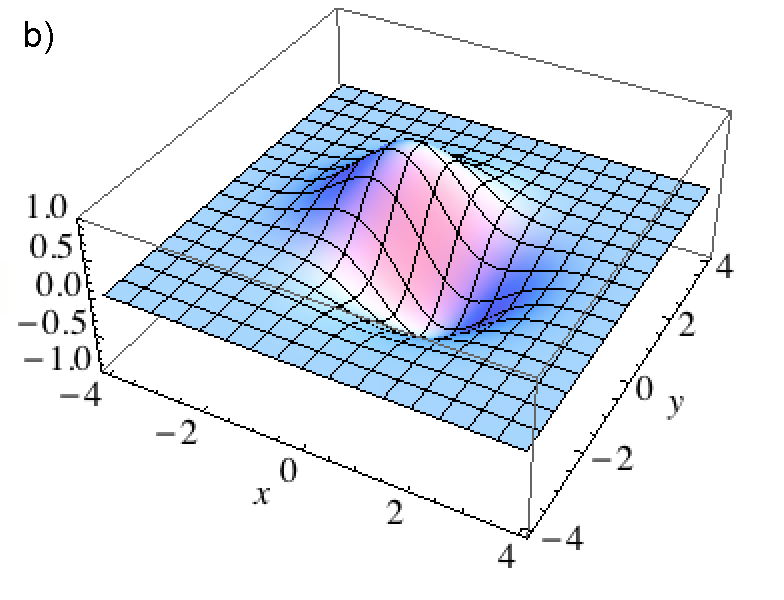}

\includegraphics[width=0.48\textwidth]{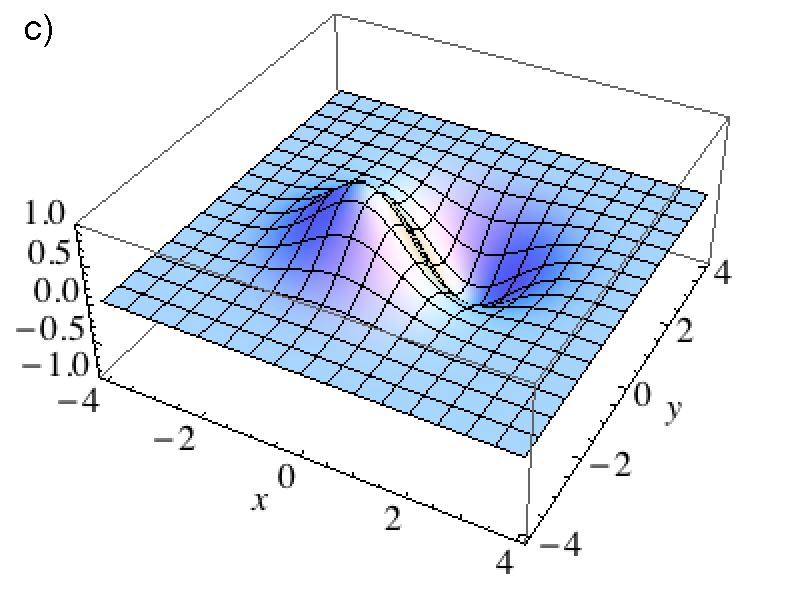}
\includegraphics[width=0.51\textwidth]{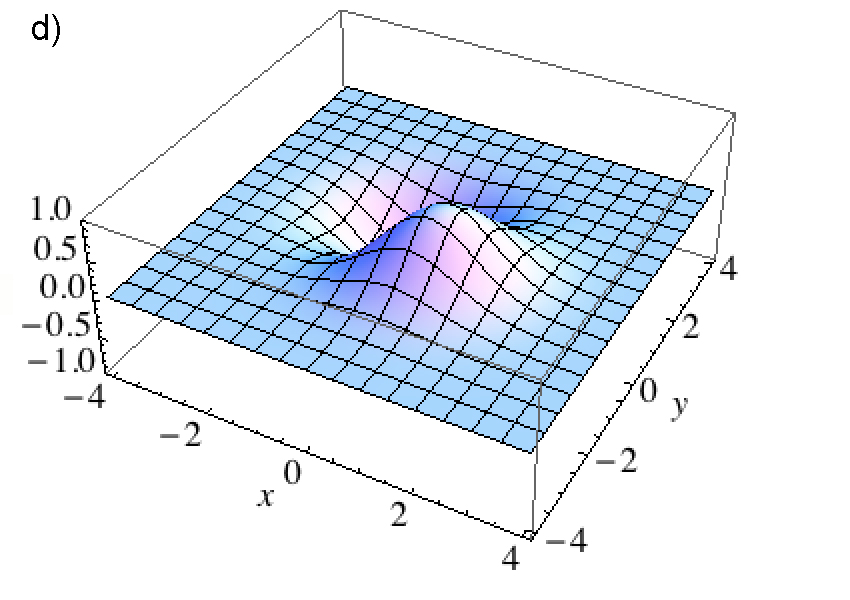}
\caption{The real part of the entangled classical wave (\ref{wave}) at four times corresponding to phase of 0, $\pi/4$, $3\pi/4$, and $5\pi/4$ radians during the period of oscillation $T = 2\pi/\tilde\omega$. The distribution rotates at constant frequency $\tilde\omega$ in the two-dimensional plane.}
\label{plots}
\end{center}
\end{figure}

The state (\ref{wave}) is a physically possible classical electromagnetic state, since each of the two terms is allowed in a two-dimensional system, and a superposition of the two is therefore also possible.  
This wavefunction is plotted in Fig.~\ref{plots} for various times.
Note that the wave function $\psi$ plotted here, which corresponds to the electromagnetic wave amplitude in our classical analog, maps to the probability wave function $\psi$ in the single-photon states (\ref{states}), while the position $x$ or $y$ here corresponds to the electromagnetic wave amplitude in the mapping (\ref{mapping}). The two spatial dimensions map to the electromagnetic wave amplitude along the two output legs of the beamsplitter discussed in the introduction.

With this state, it is manifest that the expectation value for 
having  both axes in a $|1\rangle$ state is
\begin{eqnarray}
\langle \Psi | a^\dagger_x a_x a^\dagger_ya_y |\Psi \rangle = \frac{1}{4} \int d{x}\int d{y} \ \psi^*  \left({x}^2 - \frac{\partial^2}{\partial {x}^2}-1 \right)\left({y}^2 - \frac{\partial^2}{\partial {y}^2}-1 \right) \psi =0.
\end{eqnarray}

{\bf Bell inequality}. The entangled nature of the system should allow violation of a Bell inequality, e.g. the CHSH inequality
\begin{eqnarray}
\langle {\cal O}^{(x)}_a {\cal O}^{(y)}_a)\rangle + \langle {\cal O}^{(x)}_a {\cal O}^{(y)}_b)\rangle + \langle {\cal O}^{(x)}_b {\cal O}^{(y)}_a)\rangle - \langle {\cal O}^{(x)}_b {\cal O}^{(y)}_b)\rangle \le 2,
\label{CHSH}
\end{eqnarray}
where we pick
\begin{eqnarray}
{\cal O}^{(x)}_a = S^{(x)}_z = \left(
\begin{array}{cc} 1 & 0 \\ 0 & -1 
\end{array} \right), 
\hspace{1cm}
{\cal O}^{(x)}_b =  S^{(x)}_x= \left(
\begin{array}{cc} 0 & 1 \\ 1 & 0 
\end{array} \right), 
\label{ops1}
\end{eqnarray}
which are spin-Pauli matrices acting on the $|1\rangle$ and $|0\rangle$ states of the $x$-axis. (Here the $x$ and $z$ subscripts have nothing to do with the $x-$ and $y-$axes of the cavity, which are indicated by the superscripts.) For the $y$-axis, we use 
\begin{eqnarray}
{\cal O}^{(y)}_a &=& -\frac{1}{\sqrt{2}}\left(S^{(y)}_z+S^{(y)}_x\right)
= \frac{1}{\sqrt{2}}\left(
\begin{array}{rr} -1 & -1 \\ -1 & 1 
\end{array} \right), 
\nonumber\\
{\cal O}^{(y)}_b &=& \frac{1}{\sqrt{2}}\left(S^{(y)}_z-S^{(y)}_x\right) = \frac{1}{\sqrt{2}}\left(
\begin{array}{cc} 1& -1 \\ -1 & -1 
\end{array} \right), 
\label{ops2}
\end{eqnarray}
which act on the $|1\rangle$ and $|0\rangle$ states of the $y$-axis. In terms of the continuous functions (\ref{states}), the $S_x$ operator is equivalent to
\begin{eqnarray}
S^{(x)}_x &=& a_x^\dagger(1-a_x^\dagger a_x) + a_x \nonumber\\
&=&  \frac{1}{2\sqrt{2}}\left({x}-\frac{\partial}{\partial {x}}\right) \left(3-{x}^2 +\frac{\partial^2}{\partial {x}^2} \right) + \frac{1}{\sqrt{2}}\left({x}+\frac{\partial}{\partial {x}}\right) \nonumber\\
&=&\frac{1}{2\sqrt{2}}\left(7x-x^3 +(x^2-1)\frac{\partial}{\partial x} +{x}\frac{\partial^2}{\partial x^2}-\frac{\partial^3}{\partial x^3}\right),
\end{eqnarray}
while the $S_z$ operator is equivalent to 
\begin{eqnarray}
S^{(x)}_z &=& 2a_x^\dagger a_x - 1 \nonumber\\
&=& {x}^2 - \frac{\partial^2}{\partial {x}^2}-2.
\end{eqnarray}

{\bf Measurement}. To measure the state of the system to see if it violates the Bell inequalities, we can in principle measure the electric field amplitude $\psi(x,y)$ everywhere in the cavity, perform the above operations on it analytically, and integrate over the plane, e.g.,
\begin{eqnarray}
\langle S^{(x)}_z S^{(y)}_z\rangle = \int dx dy \ \psi^*(x,y) \left( {x}^2 - \frac{\partial^2}{\partial {x}^2}-2 \right)\left( {y}^2 - \frac{\partial^2}{\partial {y}^2}-2 \right) \  \psi(x,y).
\label{integral}
\end{eqnarray}
The electric field can be measured by a set of small linear detectors adjacent to the cavity, namely polarized antennas connected to tank circuits resonant at the cavity frequency $\omega_0$. This is hard to do in the optical frequency range, but is easy to implement linear detection for electromagnetic fields in the microwave range \cite{microlinear}. Since we have assumed in all of the above that there is only one polarization of interest in the cavity, all the antennas will point in the same direction (though the orthogonal polarization could also be used to give four degrees of freedom, namely the two spatial coordinates and the two polarizations). 

For the choice of operators (\ref{ops1}) and (\ref{ops2}), this type of measurement gives a violation of the CHSH inequality, with the left side of (\ref{CHSH}) equal to $2\sqrt{2}$, as expected since we have mapped the system one-to-one to the quantum system. It is not actually necessary to measure the electric field amplitude everywhere in the plane. A violation of the Bell inequality can be obtained for a reasonable sampling of the electric field at different sites in the plane, giving a good approximation of integrals of the form (\ref{integral}).

The antenna detection implies a loss of the energy of the cavity, presumably through the cavity mirrors leaking radiation. To keep the wave function normalized, energy must be pumped into the system, as in any optical cavity. One possibility with this system is to drive the antenna detector array to pump energy into the system. In this case, positive feedback could be used to ``collapse'' the system into one or the other of the $|0\rangle$ or $|1\rangle$ states. Since the $|0\rangle$ state has positive parity and the $|1\rangle$ state has negative parity, the antennas could be set to reinforce the parity they detect. 

{\bf Conclusions}. The existence of this analog for quantum systems can help us to identify what is truly quantum and what is simply a consequence of the wave nature of quantum systems, in common with all wave systems. As we have seen here, the existence of entanglement {\em per se} is not uniquely quantum, nor are the violations of Bell inequalities which follow for entangled states. The main difference is that quantum systems can have many more possible degrees of entanglement. In quantum mechanics, each degree of freedom corresponds to a new dimension, i.e. a new orthogonal Hilbert space, with no upper limit to the number of dimensions. In classical mechanics, the number of entangled degrees of freedom is limited by the number of spatial dimensions, in a three-dimensional universe. Perhaps more than the two entangled degrees of freedom considered here can be produced in a classical system, but this analysis indicates that there is a finite upper bound of the number of entangled degrees of freedom in classical systems.
 
The CHSH inequality and other Bell's inequalities are derived for classical ``objects'' with finite countability. Such inequalities are not universal statements for all classical systems; rather they are applicable to classical systems with discrete, countable objects. In the context of classical waves, violation of a Bell inequality is not surprising. When the Bell inequalities are mapped to quantum systems, it is assumed that quantum systems also count ``objects'' which we call particles. But if we keep in mind only the continuous quantum wave functions, the violation of the Bell inequalities is no more surprising than in a classical wave. What is different in the quantum systems is that we normally think in terms of ``collapse'' of the wave functions to count a finite set of particles. In the classical analog discussed here, collapse can be forced, but it need not be. 
 
There are no nonlocal interactions in this classical system-- both orthogonal degrees of freedom exist in the same cavity. However, since they correspond to orthogonal spatial dimensions, they should be noninteracting. The spatial separation of entangled degrees of freedom appears to be a unique feature of quantum mechanical systems.

The existence of this classical analog does not fundamentally change any of the paradoxes of quantum mechanics, but it should lead us to re-examine the definitions of terms such as ``entanglement'' and what are ``truly'' quantum effects.

{\bf Acknowledgements}. This work has been supported by the National Science Foundation under grant PHY-1205762. I thank Steve Girvin and Andrew Daley for helpful conversations.


\begin{thebibliography}{99}

\bibitem{mandel} L. Mandel and E. Wolf, {\em Optical Coherence and Quantum Optics}, (Cambridge University Press, 1995), section 12.12.

\bibitem{snokebook} D.W. Snoke, {\em Solid State Physics: Essential Concepts}, (Pearson/Addison-Wesley, 2009).

\bibitem{louisell} W.H. Louisell, {\em Quantum Statistical Properties of Radiation}, (Wiley, 1973).

\bibitem{yariv} A. Yariv, {\em Quantum Electronics}, 3rd edition (Wiley, 1989).

\bibitem{botao} B. Zhang, D.W. Snoke, and A.P. Heberle, Optics Comm. {\bf 285}, 4117
 (2012).

\bibitem{heberle} R. Gordon, A.P. Heberle, and J.R.A. Cleaver, Appl. Phys. Lett. {\bf 81}, 4523 (2002).

\bibitem{photonBEC} J. Klaers, J. Schmitt, F. Vewinger, and M. Weitz, Nature {\bf 468}, 545 (2010).

\bibitem{snoke-chapter} D.W. Snoke, in {\em Exciton Polaritons in Microcavities} (Springer Series in Solid State Sciences {\bf 172}), V. Timofeev and D. Sanvitto, eds., (Springer, 2012), p. 307. 

\bibitem{polref} I. Carusotto and C. Ciuti, Rev. Modern Phys. {\bf 85}, 299 (2013).

\bibitem{berloff} M.O. Borgh, J. Keeling, and N.G. Berloff, Phys. Rev. B {\bf 81}, 235302 (2010).

\bibitem{microlinear} C. Lang et al., Nature Phys. {\bf 9}, 345 (2013).


\end{thebibliography}
\end{document}